\documentclass[doublecol,figure]{epl2}

\title{Thermodynamic reversibility in feedback processes}

\author{Jordan M. Horowitz \and Juan M.~R.~Parrondo}

\shortauthor{J.M. Horowitz and J.M.R. Parrondo}

\institute{
   Departamento de F\'isica At\'omica, Molecular y Nuclear and
   GISC, Universidad Complutense de Madrid - Universidad Complutense
   de Madrid, 28040 Madrid, Spain
  }
\pacs{05.70.-a}{Thermodynamics}
\pacs{05.20.-y}{Classical statistical mechanics}
\pacs{89.70.-a}{Information and communication theory}

\abstract{
The sum of the average work dissipated plus the information
gained  during a thermodynamic process with discrete feedback must
exceed zero. We demonstrate that the minimum  value of zero is
attained only by feedback-reversible processes that are
indistinguishable from their time-reversal, thereby extending the notion of
thermodynamic reversibility to feedback processes. In addition, we
prove that in every realization of a feedback-reversible process the
sum of the work dissipated and change in uncertainty is zero.
}

\begin{document}

\maketitle

Investigations into the thermodynamic implications of feedback have a long history \cite{Maruyama2009, Leff}, especially  with
regard to the relationship between information acquisition
and thermodynamic quantities -- such as work, heat, and
entropy. Still, there is no complete theoretical framework detailing
the relationship between feedback and thermodynamics.
Interest in developing such a framework -- a \emph{thermodynamics of feedback} -- has grown recently in part due to experiments on feedback cooling \cite{Cohadon1999, Liang2000} and feedback control of nanoparticles \cite{Cohen2005, Cohen2006}; experimental, computational, and theoretical studies of feedback driven Brownian ratchets \cite{Cao2004, Dinis2005, Feito2007, Lopez2008, Craig2008, Cao2009b}; and new theoretical predictions relating dissipation to  information \cite{Kim2004, Kim2007,
Allahverdyan2008, Sagawa2009, Ponmurugan2010, Sagawa2008, Sagawa2010, Horowitz2010,Suzuki2010, Kawai2007, Parrondo2009}.

Recently, Sagawa and Ueda derived a generalization of the second law
of thermodynamics  for quantum and classical systems manipulated by
one feedback loop \cite{Sagawa2008,Sagawa2010}, which subsequently
has been verified experimentally \cite{Toyabe2010} and extended to
classical systems driven by repeated discrete feedback  --
implemented through a  series of feedback loops initiated at
predetermined times -- independently by Horowitz and Vaikuntanathan
\cite{Horowitz2010}, and Fujitani and Suzuki \cite{Suzuki2010}.
\emph{The second law of thermodynamics for discrete feedback} states
that the average work dissipated $\langle W_{\rm d}\rangle$ in
driving a system with a discrete feedback protocol from one
equilibrium state at inverse temperature $\beta$ to another at the
same temperature is related to the microscopic information gained
through measurements $\langle I\rangle$ by
\begin{equation}\label{eq:GenSecLaw}
\beta\langle W_{\rm d}\rangle + \langle I\rangle \ge 0.
\end{equation}
Here, $\langle I\rangle$ is the mutual information between the state
of the system and the measurement  outcome
\cite{Sagawa2008,Sagawa2010,Cover}, and the average work dissipated
is the average work $\langle W\rangle$ done  in excess of the
average free-energy difference $\langle \Delta F\rangle$: $\langle
W_{\rm d}\rangle=\langle W\rangle -\langle \Delta F\rangle$. The
nomenclature derives from the observation that in the absence of
feedback [$\langle I\rangle=0$],  eq.~(\ref{eq:GenSecLaw}) reduces
to a statement of the second law of thermodynamics: $\langle W_{\rm
d}\rangle\ge 0$.

The second law of the thermodynamics plays a central role within the
framework of thermodynamics. Besides restricting the realizability
of thermodynamic processes, it distinguishes particular processes
that produce no entropy. In macroscopic systems, such processes are
called reversible, because the sequence of equilibrium states
visited by the system during the process can be traversed both
forwards and backwards \cite{Callen}. At the microscopic level the
connection between entropy production and reversibility must be
interpreted statistically and is quantified mathematically by the
distingiushibility of a thermodynamic process from its time-reversal
\cite{Kawai2007, Parrondo2009, Horowitz2009}. Like the second law of
thermodynamics, eq.~(\ref{eq:GenSecLaw}) singles out particular
processes that have no dissipation [$\beta\langle W_{\rm
d}\rangle+\langle I \rangle=0$]. Because such processes are
important in thermodynamics, it is of value to characterize them in
the presence of feedback; especially since they are optimal
processes that most efficiently convert information into work. Thus,
in this letter we analyze those processes that saturate the bound in
eq.~(\ref{eq:GenSecLaw}) [$\beta\langle W_{\rm d}\rangle+\langle I
\rangle=0$], and we demonstrate that such processes satisfy a new,
distinct criterion, similar to thermodynamic reversibility, which we
call \emph{feedback reversibility}. Intuition from the second law of
thermodynamics might lead one to naively expect that the bound in
eq.~(\ref{eq:GenSecLaw}) can always be reached as long as the
process is sufficiently slow; however, this is generally not true,
as we will demonstrate. To saturate eq.~(\ref{eq:GenSecLaw}) one
must balance $\langle W_{\rm d}\rangle$ and $\langle I\rangle$,
which requires adjusting the driving protocol, the type of
measurement, and the measurement error.

\begin{figure}
\centering
\onefigure[scale=.31]{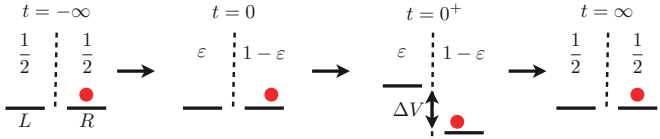}
\caption{Illustration of the feedback protocol driving  a particle (dot) that makes thermally activated transitions
between two states $L$ and $R$.
From left to right is the evolution of the relative heights of the energy levels of the two states (horizontal bars)
at four times, $t=-\infty,0,0^+$ and $\infty$, during the protocol  $\Lambda^R=\{\lambda^R_t\}_{t=0}^\infty$
associated to measuring the particle to be in state $R$ at $t=0$.
Above each state is the probability to be in that state conditioned on implementing protocol $\Lambda^R$, $\rho_t(k|\Lambda^R)$, $k=L,R$.
}
\label{fig:example}
\end{figure}

Before considering the most general situation, let us first analyze
a simple toy model of feedback, inspired by the Szilard engine
\cite{Maruyama2009,Leff}. We will demonstrate that by adjusting the
measurement error we can saturate eq.~(\ref{eq:GenSecLaw}) even when
the process is not adiabatically slow. Consider a particle weakly
coupled to a thermal bath at inverse temperature $\beta=1$ that
makes thermally  activated jumps between two states, labeled left
($L$) and right ($R$). The state energies $E_i(\lambda)$, $i=L,R$,
depend on a vector of external parameters $\lambda$ which we vary
using feedback during  a time interval $t\in(-\infty,\infty)$.
Initially at $t=-\infty$ the particle is in equilibrium with the
energies equal. From $t=-\infty$ to $0$, the system evolves freely.
At $t=0$, we measure the particle's state, misidentifying it with
error $\varepsilon$, \emph{i.e.}, the probability to measure the
particle in $R$ ($L$) given it is in $L$ ($R$)  is
$P(R|L)=\varepsilon$ [$P(L|R)=\varepsilon$]. Immediately after the
measurement at $t=0^+$, we vary the external parameters according to
a protocol that depends on the measurement outcome: if the particle
is measured to be in state $R$ ($L$), we initiate protocol
$\Lambda^R=\{\lambda^R_t\}_{t=0}^{\infty}$ ($\Lambda^L$), depicted
in fig.~\ref{fig:example},
 by  instantaneously  lowering the right (left) energy level by $\Delta V/2$,
 while simultaneously raising the left  (right) energy level by $\Delta V/2$;
 so  that the energy difference between the two states is $\Delta V$.
Finally, from $t=0^+$ to $\infty$, we quasi-statically return the energy levels to their original values.
(Similar protocols were utilized to study dissipation in nonselective quantum measurement \cite{Erez2010} and the achievability of the equality in eq.~(\ref{eq:GenSecLaw}) for quantum  feedback given a particular measurement protocol \cite{Jacobs2009}.)

Figure~\ref{fig:example} contains a depiction of a realization of
this process where the protocol $\Lambda^R$ is executed. At the four
times $t=-\infty$, $0$, $0^+$, and $\infty$ along the process, we
depict the relative heights of the energy levels of the two states.
Above each energy level is the probability at time $t$ to be in that
state \emph{conditioned} on implementing protocol $\Lambda^R$,
$\rho_t(k|\Lambda^R)$, $k=L,R$. Initially the particle is in
equilibrium with energy levels equal, therefore
$\rho_{-\infty}(R|\Lambda^R)=\rho_{-\infty}(L|\Lambda^R)=1/2$. At
$t=0$, we measure the particle to be in state $R$; consequently,
$\rho_0(R|\Lambda^R)=1-\varepsilon$ and
$\rho_0(L|\Lambda^R)=\varepsilon$ reflecting that with probability
$\varepsilon$ protocol $\Lambda^R$ is mistakenly implemented when
the particle is in state $L$. Immediately after the measurement we
instantaneously change the energy levels. Since this step is
infinitely fast, the conditional probability distributions do not
vary. Finally, starting at $t=0^+$ we infinitely slowly return the
energy levels to their original configuration, so that at $t=\infty$
the system has returned to its initial equilibrium.

\begin{figure}[htb]
\centering
\onefigure[scale=.31]{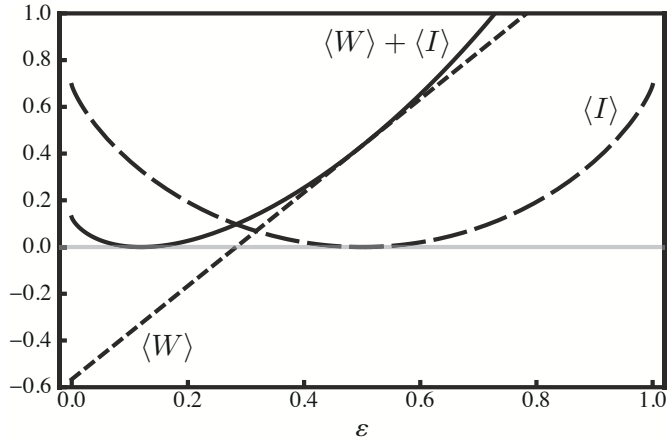}
\caption{Plot of $\langle W\rangle$ (dashed), $\langle I\rangle$ (long dashed), and their sum $\langle W\rangle +\langle I\rangle$ (solid) as a function of the measurement error $\varepsilon$ for $\Delta V=2$.}
\label{fig:exPlot}
\end{figure}

To explore how eq.~(\ref{eq:GenSecLaw}) depends on the error
$\varepsilon$, we determine the values of $\langle W_{\rm d}\rangle$
and $\langle I\rangle$ as functions of $\varepsilon$. First, observe
that since each protocol is 
cyclic the average free-energy difference is zero
$\langle \Delta F\rangle=0$, and the dissipated work equals the
work, $\langle W_{\rm d}\rangle=\langle W\rangle$. Furthermore, the
symmetry of $\Lambda^R$ and $\Lambda^L$ implies that the average
work during each protocol is the same. Thus, we focus on the average
work during $\Lambda^R$, which we calculate in two steps. First, we
compute the average work conditioned on implementing protocol
$\Lambda^R$ during the instantaneous switching of the energy levels
at $t=0^+$:
\begin{equation}\label{eq:work1}
\langle w_1\rangle_R=\varepsilon \frac{\Delta V}{2}-(1-\varepsilon)\frac{\Delta V}{2}.
\end{equation}
Second, from
time $t=0^+$ to $\infty$ the process is quasi-static.
Therefore, the
average work given $\Lambda^R$ during this period is the free-energy
difference between the equilibrium states when the energy levels differ by $\Delta V$ and when the energy levels are equal:
\begin{equation}\label{eq:work2}
\langle w_2\rangle _R=-\ln\frac{2}{e^{-\Delta V/2}+e^{\Delta V/2}}.
\end{equation}
 Noting that each protocol, $\Lambda^R$ and $\Lambda^L$, occurs with equal probability and that the work during each protocol is the same, we conclude that the total average work equals the sum of eqs.~(\ref{eq:work1}) and
(\ref{eq:work2}), which, after some algebraic manipulation, can be expressed as
\begin{equation}\label{eq:workEx}
\langle W\rangle = -\ln2-(1-\varepsilon)\ln \frac{e^{\Delta V}}{1+e^{\Delta V}}-\varepsilon\ln\frac{1}{1+e^{\Delta V}}.
\end{equation}
The mutual information $\langle I\rangle$ [see eq.~(\ref{eq:I})
below] quantifies the reduction in our uncertainty about the microscopic state of the system upon making a measurement.
It is defined as the relative entropy between the joint probability distribution of the state of the system $k=L,R$ at the time of measurement ($t=0$) and the measurement outcome $m=L,R$,
\begin{equation}
\rho_0(k,m)=
\left\{
\begin{array}
{cc}\frac{1}{2} (1-\varepsilon)& k=m \\
\frac{1}{2}\varepsilon & k\neq m
\end{array}
\right.,
\end{equation}
with the product of their respective marginal distributions, $\rho_0(k)=1/2$ and $P(m)=1/2$~\cite{Cover}.
Therefore, the mutual
information reads:
\begin{eqnarray}\label{eq:IEx}
\langle I\rangle &=& \sum_{k,m} \rho_0(k,m)\ln\frac{\rho_0(k,m)}{\rho_{0}(k)P(m)}\nonumber \\
&=& (1-\varepsilon)
\ln\left[ \frac{(1-\varepsilon)/2}{1/4}\right]+\varepsilon\ln\left[\frac{\varepsilon/2}{1/4}\right]\nonumber \\
&=&
 \ln 2+(1-\varepsilon)\ln(1-\varepsilon)+\varepsilon\ln\varepsilon.
\end{eqnarray}
In fig.~\ref{fig:exPlot} we plot $\langle W\rangle$
[eq.~(\ref{eq:workEx})], $\langle I\rangle$ [eq.~(\ref{eq:IEx})],
and their sum
\begin{equation}\label{eq:sum}
\langle W\rangle +\langle I\rangle=(1-\varepsilon)\ln \frac{(1-\varepsilon)(1+e^{\Delta V})}{e^{\Delta V}}+\varepsilon\ln\left[\varepsilon(1+e^{\Delta V})\right]
\end{equation}
 as functions of
the error $\varepsilon$.
For error-free measurements ($\varepsilon=0$), we are able to
extract the maximum amount of work [$-\langle W\rangle$ is largest]
and obtain the maximum amount of information $\langle I\rangle =\ln
2$. However, $\langle W\rangle +\langle I \rangle >0$; some of the
information is not used to extract work. At the expense of
increasing $\varepsilon$, decreasing the amount of work extracted,
and decreasing the amount of information gained, we can reach the
bound in eq.~(\ref{eq:GenSecLaw}). From eq.~(\ref{eq:sum}), we see
that the sum $\langle W\rangle +\langle I\rangle$ is zero when
$\varepsilon$ equals
\begin{equation}\label{eq:epRev}
\varepsilon_0=\frac{1}{1+e^{\Delta V}}.
\end{equation}
Thus, by adjusting the measurement error with fixed external
parameter protocols we can construct a feedback protocol that is not
adiabatically slow and satisfies $\beta\langle W_{\rm d}\rangle
+\langle I\rangle = 0$.

\begin{figure}[thb]
\centering
\onefigure[scale=.31]{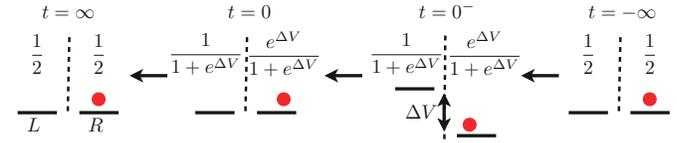}
\caption{Illustration of the reverse process at four times, $t=-\infty,0^-,0$ and $\infty$ (right to left), along the protocol  $\tilde\Lambda^R$.
Above each state is the conditional probability distribution  $\tilde\rho_t(k|\tilde\Lambda^R)$, $k=L,R$.
}
\label{fig:example2}
\end{figure}

A physical interpretation can be given to $\varepsilon_0$ in
eq.~(\ref{eq:epRev}) by considering the time-reversal of the
preceding feedback process. The time-reversal of a feedback process
is not trivial, since there is no such thing as the time-reversal of
a measurement. However, we follow Ref.~\cite{Horowitz2010} and
construct a distinct thermodynamic process termed the \emph{reverse}
process, which will act as the time-reversed feedback process. The
reverse process begins by randomly selecting a protocol, $\Lambda^L$
or $\Lambda^R$, according to the probability  that the respective
measurement outcome, $L$ or $R$, occurs during feedback:
$P(L)=P(R)=1/2$. The particle is then driven away from equilibrium
by using the time-reversal of the selected protocol,
$\tilde\Lambda^k=\{\lambda_{-t}^k\}_{t=-\infty}^{\infty}$, $k=L,R$. Now,
imagine randomly selecting protocol $\Lambda^R$, and consider the
evolution of the system conditioned on executing the time-reversed
protocol $\tilde\Lambda^R$,  depicted in fig.~\ref{fig:example2}.
Initially the particle is in equilibrium with energies equal. As a
result,  the probabilities to find the particle in states $L$ and
$R$ at $t=-\infty$ in the reverse  process \emph{conditioned} on the
protocol $\tilde\Lambda^R$ are initially equal:
$\tilde\rho_{-\infty}(R|\tilde\Lambda^R)=\tilde\rho_{-\infty}(L|\tilde\Lambda^R)=1/2$.
From $t=-\infty$ to $0^-$,  the right energy level is
quasi-statically lowered by $\Delta V/2$ while the left energy level
is raised by $\Delta V/2$. Thus, at $t=0^-$ the conditional
probability distribution is  a Boltzmann distribution:
$\tilde\rho_{0^-}(L|\tilde\Lambda^R)=1/(1+e^{\Delta V})$ and
$\tilde\rho_{0^-}(R|\tilde\Lambda^R)=e^{\Delta V}/(1+e^{\Delta V})$.
At $t=0$, the energy levels are instantaneously returned to their
original equal values; the conditional probablity distribution does
not change. Finally, from $t=0$ to $\infty$ the energy levels are
held fixed, while the system relaxes back to its initial
equilibrium.

Comparing figs.~\ref{fig:example} and \ref{fig:example2}, we see
that when $\varepsilon=\varepsilon_0$ [eq.~(\ref{eq:epRev})] the
conditional probability distributions in the feedback and the reverse
processes are equal at the main stages of the process:
\begin{equation}\label{eq:rev_conditional}
\rho_t(k|\Lambda^R)=\tilde\rho_{-t}(k|\tilde\Lambda^R),
\end{equation}
for $k=L,R$.
The same holds for $\Lambda^L$.
 Below we demonstrate that this equality holds at any time $t$ [see eq.~(\ref{eq:rev2})].
 Moreover, since each protocol is implemented with equal likelihood
in both the original feedback process and the reverse process, the
joint distributions of states and protocols for the feedback and
reverse processes are equal for an optimal feedback process:
$\rho_t(k,\Lambda^m)=\tilde\rho_{-t}(k,\tilde\Lambda^m)$, $k,m=L,R$.
Notice that the reverse and feedback processes are of a different
nature: there are no measurements in the reverse process. Despite
this difference, identifying the reverse processes as the
time-reversed feedback processes is consistent with the
thermodynamic principle linking reversibility and dissipation: the
process with least dissipation -- the protocol which saturates
eq.~(\ref{eq:GenSecLaw}) -- corresponds to the situation where the
feedback process is indistinguishable from its time-reversal (from
the reverse process). Consequently, we call these processes
\emph{feedback reversible} or simply \emph{reversible}.

We now demonstrate that this conclusion holds quite generally as a
consequence of the detailed fluctuation theorem for discrete
feedback [eq.~(\ref{eq:dfr}) below]~\cite{Horowitz2010}.
This theorem relates the fluctuations in two thermodynamic processes related by time-reversal: the forward and the reverse processes.
Our analysis begins by sketching a derivation of
eq.~(\ref{eq:GenSecLaw}) based on the detailed fluctuation theorem
for discrete feedback originally
presented in Ref.~\cite{Horowitz2010}. For clarity of exposition, we
limit our discussion to feedback processes with only one feedback
loop initiated at the beginning of the process. All our conclusions
can be generalized to situations with repeated discrete feedback.

Consider a classical system, whose position in phase space (or
microscopic configuration) at time $t$ is $z_t$. This system is
driven away from equilibrium at inverse temperature $\beta$ from
$t=0$ to $\tau$ using one feedback loop as follows: at $t=0$ a
physical observable $M$ is measured with outcomes $m$  that occur
with probability $P(m|z_0)$, conditioned on the system's state at
the time of measurement $z_0$. Based on the outcome of this
measurement, the system is driven by varying a set of external
parameters $\lambda$ with time using the protocol $\Lambda^m=\{
\lambda_t^m\}_{t=0}^\tau$ from $\lambda_0^m=A$ to
$\lambda_\tau^m=B^m$. Finally, at time $t=\tau$ the external
parameters are held fixed at $\lambda^m_\tau=B^m$ while the system
relaxes back to equilibrium at inverse temperature $\beta$.
Repeatedly executing this sequence of actions -- each time
equilibrating the system, driving the system using feedback, and
then re-equilibrating the system -- generates an ensemble of
realizations of  the \emph{forward} process. In each realization,
thermal fluctuations will cause the system to trace out a different
microscopic trajectory through phase space
$\gamma=\{z_t\}_{t=0}^\tau$. The joint probability to observe
$\gamma$ with $\Lambda^m$ is $\mathcal{P}[\gamma;\Lambda^m$]. The
work dissipated  during this realization, 
\begin{equation}
W_{\rm d}[\gamma;\Lambda^m]=W[\gamma;\Lambda^m]-\Delta F[\Lambda^m],
\end{equation}
is the work $W[\gamma;\Lambda^m]$ done in excess of the free-energy difference $\Delta F[\Lambda^m]$. Here, $\Delta F[\Lambda^m]$ depends on the executed protocol, because the final external parameter value is a function of the measurement, $\lambda^m_\tau=B^m$.
Moreover, measurements made upon initiating the feedback loop change
our uncertainty about the microscopic state of the system by an
amount \cite{Horowitz2010}
\begin{equation}\label{eq:I}
I[\gamma;\Lambda^m]=\ln\left[\frac{P(m|z_0)}{P(m)}\right],
\end{equation}
where $P(m)$ is the (unconditional) probability to observe outcome $m$ when measuring the physical observable $M$.
The average of $I$ over many realizations is the mutual information that appears in eq.~(\ref{eq:GenSecLaw}).

The reverse process is defined as in our previous example. In the
reverse process no measurements are made. Instead, we drive the
system away from equilibrium using a \emph{reverse} protocol
$\tilde\Lambda^m=\{\tilde\lambda_{t}^m\}_{t=0}^\tau$ with
$\tilde\lambda_t^m=\lambda_{\tau-t}^m$, which is selected randomly
with probability $\tilde\pi[\tilde\Lambda^m]$ defined to be equal to
the probability to implement the forward protocol
$\lambda^m_{t}=\tilde\lambda^m_{\tau-t}$ in the forward process
$\pi[\Lambda^m]$:
\begin{equation}\label{eq:pi}
\tilde\pi[\tilde\Lambda^m]=\pi[\Lambda^m]=\int d\gamma\, \mathcal{P}[\gamma;\Lambda^m],
\end{equation}
where $d\gamma$ is a measure on microscopic trajectory space. After
selecting $\tilde\Lambda^m$, the system is equilibrated with a
thermal bath at  inverse temperature $\beta$ with external
parameters fixed at $\tilde\lambda^m_0=\lambda^m_\tau=B^m$.
Observe that the initial equilibrium distribution of the reverse process corresponds to external parameter value $B^m$ and depends on which protocol $\tilde\Lambda^m_t$ is implemented \cite{Horowitz2010}.
From $t=0$ to $\tau$ the external parameters are varied according to
the reverse protocol
$\tilde\Lambda^m=\{\tilde\lambda^m_t\}_{t=0}^\tau$. At $t=\tau$, the
external parameters are fixed to $\tilde\lambda^m_\tau=\lambda_0=A$
while the system relaxes back to equilibrium. For every microscopic
trajectory of the forward process $\gamma=\{z_t\}_{t=0}^\tau$, there
is a \emph{conjugate} reverse trajectory
$\tilde\gamma=\{\tilde{z}_{t}\}_{t=0}^\tau$, where
$\tilde{z}_t=z^*_{\tau-t}$ and $z^*$ denotes momentum reversal. The
probability to observe reverse trajectory $\tilde\gamma$ and reverse
protocol $\tilde\Lambda^m$ in the reverse process is
$\tilde{\mathcal P}[\tilde\gamma;\tilde\Lambda^m]$.

Having introduced notation and presented the definitions of the forward and reverse processes, we now state the detailed fluctuation theorem \cite{Horowitz2010}:
\begin{equation}\label{eq:dfr}
\frac{\mathcal {P}[\gamma;\Lambda^m]}{\tilde{\mathcal P}[\tilde\gamma;\tilde\Lambda^m]}=e^{\beta W_{\rm d}[\gamma;\Lambda^m] +I[\gamma; \Lambda^m]}.
\end{equation}
Because no measurements are made in the reverse process, there are microscopic trajectories and reverse protocols that occur together in the reverse process whose conjugate microscopic trajectories and conjugate protocols do \emph{not} occur together in the forward process; it is possible for $\tilde{\mathcal P}\neq0$ when $\mathcal{P}=0$ \cite{Horowitz2010}.

Equation~(\ref{eq:dfr}) implies that the relative entropy,
$D(f||g)=\int dx f(x)\ln(f(x)/g(x))$, between $\mathcal{P}$ and
$\tilde{\mathcal P}$ is~\cite{Horowitz2010}
\begin{equation}\label{eq:relEntFluc}
D(\mathcal{P}||\tilde{\mathcal P} )= \beta\langle W_{\rm d}\rangle +\langle I\rangle,
\end{equation}
where $\langle\cdot\rangle$ is an ensemble average over realizations of the forward process.
$D(\mathcal{P}||\tilde{\mathcal P}$) measures the distinguishability of the forward and reverse processes; it is a microscopic measure of the intensity of the ``arrow of time'' \cite{Kawai2007,Parrondo2009}.

Equation (\ref{eq:GenSecLaw}) now follows by exploiting the
nonnegativity of the relative entropy ($D\ge 0$) \cite{Cover} in
eq.~(\ref{eq:relEntFluc}). Moreover, eq.~(\ref{eq:relEntFluc})
implies that thermodynamic processes for which $\beta\langle W_{\rm
d}\rangle +\langle I\rangle=0$ are those with $D=0$, which is true
if and only if
\begin{equation}\label{eq:rev1}
\mathcal{P}[\gamma;\Lambda^m]=\tilde{\mathcal P}[\tilde\gamma;\tilde\Lambda^m]
\end{equation}
for all $\gamma$ and $\Lambda^m$ \cite{Cover}. The condition $D=0$
additionally requires that the supports of $\mathcal{P}$ and
$\tilde{\mathcal P}$  -- the sets of microscopic trajectories and
protocols for which $\mathcal{P}$ and $\tilde{\mathcal P}$ are
nonzero -- must be identical: every microscopic trajectory and
reverse protocol that occur together in the reverse process have
conjugate pairs that occur together in the forward process.
Equation~(\ref{eq:rev1}) is a microscopic statement of
reversibility: the process looks the same forwards and backwards in
time, since every realization occurs with equal likelihood in the
forward and reverse processes. We conclude that the inequality in
eq.~(\ref{eq:GenSecLaw}) is saturated only when eq.~(\ref{eq:rev1})
is satisfied,  that is \emph{only for reversible processes}.

One could in principle consider a ``super-system''
composed of our system of interest and a feedback mechanism --  formed from a controller that manipulates the parameters $\lambda$ and a memory that records the measurement outcomes. 
While the explicit implementation of such a set-up with a time-dependent Hamiltonian is a difficult task, it is still instructive to assume that a super-system exists and to compare the feedback reversibility 
we have introduced with standard thermodynamic reversibility. 
In this set-up, we manipulate the super-system by varying a collection of parameters external to both the system of interest and the feedback mechanism according to a predetermined cyclic protocol.
The work performed by the external parameters along this cycle is responsible for recording the measurement of the system, the control in response to the measurement, and the erasure of the measurement.
As shown by Sagawa and Ueda \cite{Sagawa2009}, under rather general
hypothesis, the work to measure $\langle W_{\rm meas}\rangle$ plus the work to erase $\langle W_{\rm eras}\rangle$ is bound by $\beta(\langle W_{\rm eras}\rangle+\langle
W_{\rm meas}\rangle )\geq \langle I\rangle$. 
Combining this bound with equation~(\ref{eq:GenSecLaw}), we find that the total work performed on the super-system is 
\begin{equation}\label{eq:supersystem} \langle W_{\rm d}\rangle+\langle W_{\rm eras}\rangle+\langle W_{\rm
meas}\rangle\geq 0,
\end{equation}
which is the second law of thermodynamics applied to the super-system when the feedback mechanism has gone through a cycle.
To achieve the equality in equation~(\ref{eq:supersystem}), and consequently in
(\ref{eq:GenSecLaw}), one would need thermodynamic reversibility in the full super-system -- that is the thermodynamic process must be indistinguishable from its time-reversal in the phase space of the super-system.
However, such a process is not quasi-static, because during the writing and erasure of the memory the super-system is not in equilibrium.
We have shown that to achieve the optimal feedback control, equality in (\ref{eq:GenSecLaw}), one only needs feedback reversibility in the system, while the controller can be out of equilibrium.

From eq.~(\ref{eq:rev1}) follows another useful microscopic
statement of reversibility in terms of the phase space densities
conditioned on the executed protocol [eq.~(\ref{eq:rev2}) below].
Recall that in our construction of the reverse process, the
probability to execute $\Lambda^m$ in the forward process
$\pi[\Lambda^m]$ equals the probability to use the conjugate reverse
protocol $\tilde\Lambda^m$ in the reverse process
$\tilde\pi[\tilde\Lambda^m]$ [eq.~(\ref{eq:pi})]. Thus dividing both
sides of eq.~(\ref{eq:rev1}) by
$\pi[\Lambda^m]=\tilde\pi[\tilde\Lambda^m]$ we find that, for a
reversible process, the statistics of the trajectories
\emph{conditioned} on the protocol executed in the forward and
reverse processes are also indistinguishable:
\begin{equation}
\mathcal{P}[\gamma|\Lambda^m]=\tilde{\mathcal P}[\tilde\gamma|\tilde\Lambda^m].
\end{equation}
Furthermore, by integrating these conditional trajectory distributions over
all trajectories that pass through
$z_t=\tilde{z}_{\tau-t}^*$, we find that the conditional phase space
densities must also be equal:
\begin{equation} \label{eq:rev2}
\rho(z_t|\Lambda^m)=\tilde\rho(\tilde{z}_{\tau-t}|\tilde\Lambda^m),
\end{equation}
which is the general expression corresponding to eq.~(\ref{eq:rev_conditional}).

Equation~(\ref{eq:rev2}) is valid at all times during a
feedback-reversible process. In particular, it implies that in our
preceding example eq.~(\ref{eq:rev_conditional}) is true even during
the time interval $t\in(-\infty,0)$ prior to the measurement. The
reversibility of the process during this interval may be surprising
at first, since during this period the system is freely evolving
with the external parameters held fixed. To gain further insight,
consider a special case of our example with no measurement error,
$\varepsilon=0$. In the forward process during the interval
$t\in(-\infty,0)$, $\rho_t(k|\Lambda^R)$ represents the evolution of
the (probabilistic) state of the system conditioned on the  {\em
future} measurement outcome being $R$ at $t=0$. Whereas during this
interval, $\tilde\rho_{-t}(k|\tilde\Lambda^R)$ is a relaxation from
a non-equilibrium distribution
$\tilde\rho_{0}(k|\tilde\Lambda^R)=\delta_{k,R}$ to equilibrium
$\tilde\rho_{\infty}(k|\tilde\Lambda^R)=1/2$. The reversibility
condition eq.~(\ref{eq:rev2}) [eq.~(\ref{eq:rev_conditional})]
states that these two processes are identical upon time reversal.
\begin{figure}[htb]
\centering
\onefigure[scale=.41]{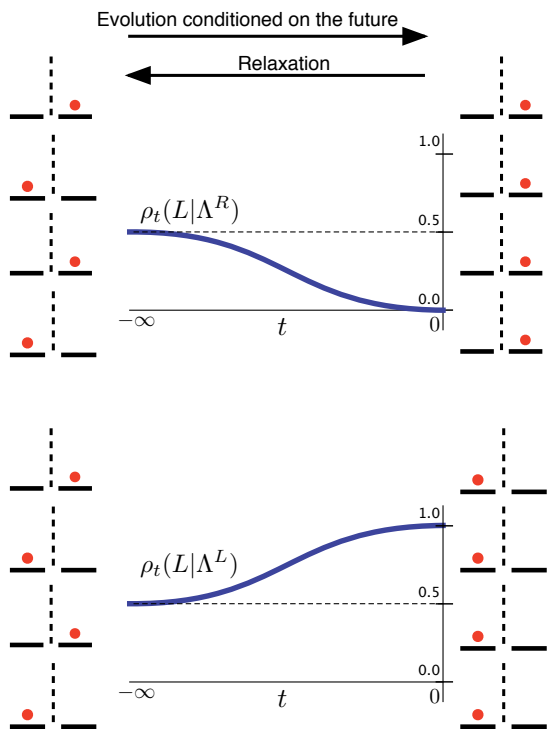}
\caption{
During the time interval $t=-\infty$ to $0$, the external parameters are constant with time; therefore, an ensemble of systems initially in equilibrium remains in equilibrium.
However, the sub-ensemble of systems that at $t=0$ are measured to be in the right (left) state -- depicted here as the four upper (lower) copies of the system -- does vary in time.
The probability in this sub-ensemble to be in state $L$ at time $t$, $\rho_{t}(L|\Lambda^{R})$ [$\rho_{t}(L|\Lambda^{L})$], is plotted as a function of time during the interval $t\in (-\infty,0)$, assuming error-free measurement ($\varepsilon=0$).
In a reversible process, the time reversal of this evolution is identical to a
relaxation to equilibrium from a nonequilibrium state given by $\rho_{0}(L|\Lambda^{R})=0$ [$\rho_{0}(L|\Lambda^{L})=1$].}
\label{fig:revrelax}
\end{figure} 
In fig.~\ref{fig:revrelax}, we illustrate this reversibility by
depicting the evolution of an ensemble of systems during
$t\in(-\infty,0)$. The ensemble is in equilibrium from $t=-\infty$
to $0$, but the sub-ensemble of systems that are in the right (left)
state at the time of measurement $t=0$ exhibits an evolution that is
identical to the time-reversal of the corresponding relaxation. The
underlying reason is microscopic reversibility:
 the probability in equilibrium to observe a microscopic trajectory and its time reversal are equal~\cite{Crooks2000}.
As a consequence, the probability to observe a microscopic trajectory conditioned on being at a particular state  in the future, say $z^\prime$,  is the same as the probability to observe the time reversal of this trajectory while the system relaxes to equilibrium when initially at $z^\prime$.

Finally,  we note that in \emph{every} realization of a reversible
process $\beta  W_{\rm d}+I=0$, which follows by substituting the
reversibility condition eq.~(\ref{eq:rev1}) into the nonequilibrium
detailed fluctuation theorem [eq.~(\ref{eq:dfr})]. Although $\beta
W_{\rm d}+I=0$ for every realization of a reversible process, the
value of $W_{\rm d}$ ($I$) may differ in each realization.

We have demonstrated that those feedback processes that most
efficiently use information to extract work [that saturate
eq.~(\ref{eq:GenSecLaw})] are feedback reversible --  they are
indistinguishable from their time-reversal, thereby extending the
concept of thermodynamic reversibility to feedback processes. Like
reversible thermodynamic processes, feedback-reversible processes
are ideal for experimentally measuring free-energy differences.
When all external parameter protocols end at the same value $\lambda^m_\tau=B$ -- independent of $m$ -- the sum of the work and information along a feedback-reversible process equals the free-energy difference.
Feedback-reversible processes can also be used to estimate the information gain by measuring the work dissipated.
In our illustrative example, if we were to measure the work as as function of $\Delta V$ its minimum value would equal $\langle I\rangle$ [eq.~(\ref{eq:workEx})].
In general, a measurement of the work disspated gives a lower bound on the information according to eq.~(\ref{eq:GenSecLaw}).
Furthermore, our
work  gives an indication on how to design feedback-reversible
protocols for a given measurement scheme, which may be a difficult
task \cite{Kim2011,Abreu2011}. The protocols must be such that
eq.~(\ref{eq:rev2}) holds at every instant. In particular,
immediately after a measurement with outcome $m$, the conditioned
phase space density must be identical to the conditioned density
prepared by the protocol $\lambda^m_t$ run in reverse. This identity
is the key ingredient for designing an optimal protocol.

\acknowledgements
The authors acknowledge financial support from Grant MOSAICO (Spanish Government) and MODELICO (Comunidad de Madrid).

\bibliographystyle{eplbib.bst}
\bibliography{Feedback,PhysicsTexts,FluctuationTheory}

\end{document}